# DISCUSSION OF: BAYESIAN VIEWS OF AN ARCHAEOLOGICAL FIND

By Joseph B. Kadane

*Carnegie Mellon University*

Andrey Feuerverger (2008) is to be congratulated on having given us such a careful analysis of a very interesting data set. He has obviously gone to great efforts to understand the archaeology (in several languages) and background of the tomb in question, and the literature and history surrounding it. This effort is exactly what a modern statistician should be doing in an applied problem.

Unfortunately Feuerverger is hampered, in my view, by his predisposition toward sampling theory. His technique relies on his RR-values ("relevance and rareness"), but he gives no theory of RR. Just what is it? What justifies multiplying them together? What have you got when you're done? Second, his method computes the probability of data as or more extreme than that observed were the null hypothesis true, which violates the likelihood principle because it counts as relevant data that did not occur. Finally, his method is very limited in the conclusions it permits one to draw: either the null hypothesis is false or something unusual has happened. Well, which is it? Using his paradigm, he is unable even to give a probability of which of these is the case. A great deal of effort goes into establishing a conclusion whose form does not address the question of interest, at least as I interpret it.

By contrast, a Bayesian treatment has clear-cut and simple rules. These have been worked out extensively for problems in forensic science; indeed the present problem can be so regarded. The question, as Feuerverger himself points out, is to calculate $P(B \mid A)/P(B \mid \overline{A})$ where $A$ is the event that the Talpiyot tomb is that of the $NT$ family, and $\overline{A}$ is that it is not. The event $B$ is the evidence we have, namely the specific names found in Talpiyot. $P(B \mid A)$ is probability of this tomb arising if it were the tomb of the $NT$ family. Thus it involves what other renditions of names might have been used for the persons in the $NT$ family, and the possible identities of the unidentified persons in the tomb. Similarly $P(B \mid \overline{A})$, which is essentially what he is computing from the onomasticon, is the probability of this configuration arising from some other family or group of people. While he says that this specification of $B$ is "awkward to work with," it seems to me that it leads us to address the essential questions in analyzing the Talpiyot tomb.

Höfling and Wasserman (2008) and Ingermanson (2008) in preceding comments on the paper give differing Bayesian analyses of this problem, and Mortera and Vicard (2008) stated how they would use DNA analysis in one. Should we be disturbed that the former two make different assumptions, and derive different posterior probabilities? I would argue not. The strength of the Bayesian approach is that it requires the assumptions to be stated explicitly and argued for. The acceptability of those assumptions is for each reader to judge for himself or herself. All the Bayesian argument ensures is that each writer is coherent, that is, does not contain internal contradictions in a certain technical sense. Thus the Bayesian view of probability is arguably like a language. That a sentence is in grammatical English does not require the reader to agree with it; proper grammar only helps us to understand what the writer means. Similarly, an opinion expressed in probabilistic terms is explicit, that is, a reader can understand what the writer's view is, but it is up to the writer to be persuasive to the reader. Each reader, then, needs to state the beliefs found most congenial, and to compute his or her own posterior probability accordingly.

Finally, it is obviously necessary to say something about how the statistical analysis of this data set relates to the religious beliefs of many people. Fortunately there is no contradiction between the Bayesian paradigm and such beliefs. Bayes Theorem in odds form reads, as Feuerverger points out,

$$\frac{P(A \mid B)}{P(\overline{A} \mid B)} = \frac{P(A)}{P(\overline{A})} \times \frac{P(B \mid A)}{P(B \mid \overline{A})}.$$

Here the factor $P(A)/P(\overline{A})$ is the prior odds of the event $A$. For those whose religious beliefs specify $P(A) = 0$ and $P(\overline{A}) = 1$ (i.e., there is no chance that the tomb is that of the $NT$ family), whatever the likelihood contribution [here $P(B \mid A)/P(B \mid \overline{A})$], the posterior odds of $A$ [here $P(A \mid B)/P(\overline{A} \mid B)$] are zero. This set of beliefs is coherent in the technical sense (i.e., it does not lead to sure loss), and hence is fully consistent with the Bayesian view.

Department of Statistics
Baker Hall 232A
Carnegie Mellon University
Pittsburgh, Pennsylvania 15213
USA
E-mail: kadane@stat.cmu.edu